\documentclass[sigconf, screen]{acmart}
\AtBeginDocument{%
  \providecommand\BibTeX{{%
    \normalfont B\kern-0.5em{\scshape i\kern-0.25em b}\kern-0.8em\TeX}}}

\setcopyright{acmlicensed}
\copyrightyear{2018}
\acmYear{2018}
\acmDOI{XXXXXXX.XXXXXXX}

\acmConference[Conference acronym 'XX]{Make sure to enter the correct
  conference title from your rights confirmation emai}{June 03--05,
  2018}{Woodstock, NY}
\acmBooktitle{Woodstock '18: ACM Symposium on Neural Gaze Detection,
 June 03--05, 2018, Woodstock, NY} 
\acmISBN{978-1-4503-XXXX-X/18/06}

\usepackage{xcolor}
\definecolor{myred}{rgb}{0.6, 0.0, 0.0}
\definecolor{myblue}{rgb}{0, 0.0, 0.8}
\newcommand{\tibin}[1]{#1}

\newcommand{\tibilm}[1]{#1}

\begin{document}



\title{DragD3D: Realistic Mesh Editing with Rigidity Control Driven by 2D Diffusion Priors}




\author{Tianhao Xie}
\affiliation{%
  \institution{Concordia University}
  \city{Montreal}
  \country{Canada}}
\email{tianhao.xie@mail.concordia.ca}

\author{Eugene Belilovsky}
\affiliation{%
 \institution{Concordia University and MILA}
 \city{Montreal}
 \country{Canada}}
\email{eugene.belilovsky@concordia.ca}
\author{Sudhir Mudur}
\affiliation{%
  \institution{Concordia University}
  \city{Montreal}
  \country{Canada}
}
\email{mudur@cse.concordia.ca}

\author{Tiberiu Popa}
\affiliation{%
  \institution{Concordia University}
  \city{Montreal}
  \country{Canada}}
\email{tiberiu.popa@concordia.ca}

\renewcommand{\shortauthors}{Trovato and Tobin, et al.}

\begin{abstract}
  Direct mesh editing and deformation are key components in the geometric modeling and animation pipeline. Mesh editing methods are typically framed as optimization problems combining user-specified vertex constraints with a regularizer that determines the position of the rest of the vertices. The choice of the regularizer is key to the realism and authenticity of the final result. Physics and geometry-based regularizers are not aware of the global context and semantics of the object, and the more recent deep learning priors are limited to a specific class of 3D object deformations. Our main contribution is a vertex-based mesh editing method called DragD3D based on (1) a novel optimization formulation that decouples the rotation and stretch components of the deformation and combines a 3D geometric regularizer with (2) the recently introduced DDS loss which scores the faithfulness of the rendered 2D image to one from a diffusion model. Thus, our deformation method achieves globally realistic shape deformation which is not restricted to any class of objects.
Our new formulation optimizes directly the transformation of the neural Jacobian field explicitly separating the rotational and stretching components.
The objective function of the optimization combines the approximate gradients of DDS and the gradients from the geometric loss to satisfy the vertex constraints. 
 Additional user control over desired global shape deformation is made possible by allowing explicit per-triangle deformation control as well as explicit separation of rotational and stretching components of the deformation. We show that our deformations can be controlled to yield realistic shape deformations that are aware of the global context of the objects, and provide better results than just using geometric regularizers.
\end{abstract}



\begin{teaserfigure}
  \includegraphics[width=\textwidth]{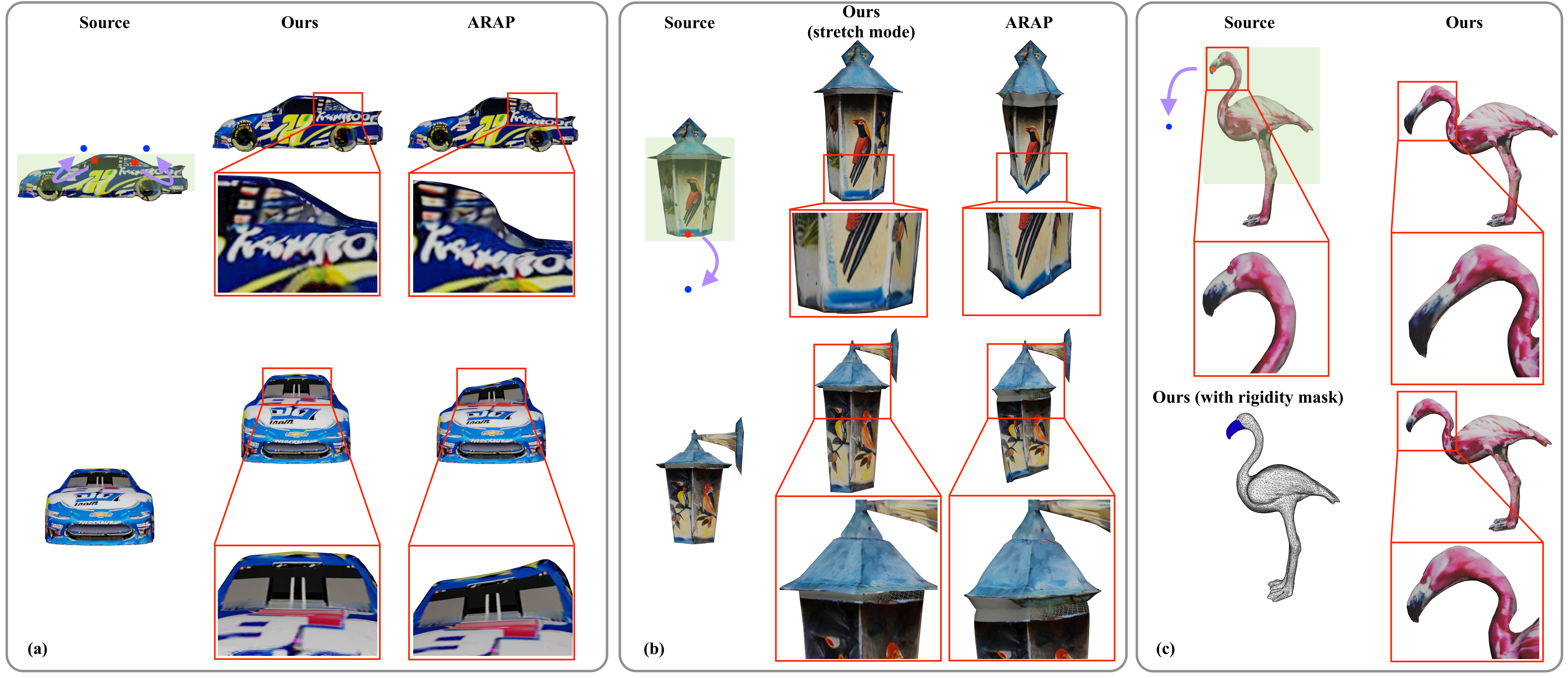}
  \caption{Our vertex-based local editing achieves realistic global context-aware mesh deformation using a few handle points. The user specifies a few mesh vertices (red disks) as handles, their target positions (blue disks), and a region of influence (light green). They can optionally specify a deformation type (e.g. stretch only $(b)$ and/or a rigidity mask $(c)$).}
  \label{fig:teaser}
\end{teaserfigure}


\maketitle

\section{Introduction}
Geometric deformation and editing are fundamental operations in the geometric modeling pipeline that have received a lot of attention over the years.
Among the many varieties of geometric representations and editing modalities, vertex-based mesh editing through direct manipulation of mesh vertices is particularly appealing for many applications and constitutes the main focus of this work. 
Classical direct mesh editing methods~\cite{botsch2007linear,yuan2021revisit} employ an optimization framework where the user vertex constraints are complemented by a regularizer whose main goal is to keep the rest of the shape realistic. 
To accomplish this, regularizers either use elasticity or geometric priors.
These regularizers try to minimize locally computed energies and they often fail for large deformations because: (1) they assume the deformation behavior is homogeneous across the object, which is not true in practice~\cite{sumner2005mesh,popa2007interactive} and (2), especially for CAD models, there are global semantic relationships between different parts that are lost when only local regularizers are considered. 
Methods such as iWires~\cite{gal2009iwires} attempted to address this issue by first analyzing the shape to extract global relationships between parts using a network of curves, but this is not a general solution.

One way to address these shortcomings is to use data-driven methods.
Data-driven methods use three main strategies: \\(1) learn the shape space of certain classes of objects and use it to guide the deformations~\cite{yumer2016learning,wang20193dn,yifan2020neural,jakab2021keypointdeformer,deng2021deformed,liu2021editing,yang2021learning,shechter2022neuralmls,hyung2023local}\\(2) learn the deformation behavior from a set of example deformations in a supervised manner~\cite{sumner2005mesh,popa2007interactive,tang2022neural}.
The biggest challenge in the above two strategies is the reliance on real 3D data. 
While the collection of 3D datasets available for research has greatly improved in the last few years, it is minuscule compared to the richness of 2D images available; it can be argued that a large gap between these two will always exist due to the inherent challenges in 3D acquisition compared to 2D acquisition.
\\(3) Another strategy is to rely on optimizing the shape using 2D priors obtained from large pre-trained 2D generative models such as CLIP~\cite{mohammad2022clip,gao2023textdeformer,sanghi2023clip} or stable diffusion~\cite{poole2022dreamfusion,mikaeili2023sked,tsalicoglou2023textmesh}, and guide the deformation through differential rendering.
\tibilm{
Although these works have demonstrated the generation of 3D objects from scratch, an open question is whether large-scale 2D diffusion models embed 3D geometry knowledge well enough to enable their use in user-controlled geometry-related tasks such as mesh manipulation. Our work is an important first step towards understanding this. 
}

\tibin{Many challenges exist related to using 2D diffusion priors obtained from large pre-trained 2D models to guide the 3D deformation.
We note three of the most fundamental challenges that have been observed:
(1) A large variety of views are necessary to guide the deformation (2) the DDS is noisy and will not be influenced by small local variations of the geometry, thus requiring a well-designed geometric regularizer and (3) the 2D prior keeps the deformed object within the class of objects specified in the text, but, as noted in DragGAN~\cite{pan2023drag}, allows changing the identity of the objects, which is undesirable for a mesh editing and deformation framework.
}

\tibilm{In this work, we propose for the first time a method that combines classical fine vertex-controlled mesh-based deformations with 2D priors from a large-scale 2D diffusion model.}
We frame the deformation as an optimization problem 
where we directly optimize the transformation of the Jacobian field represented as a combination of rotation and stretching. This parameterization of the deformation problem allows us to address the last two of the challenges mentioned above.
\tibin{We formulate a geometric loss to keep the deformation smooth and we combine it with a vertex constraints loss and a 2D diffusion score obtained from multiple renderings of the object scored against large-scale 2D priors to provide the global context. Geometry changes are restricted to the region of influence.
To address identity preservation, we incorporate an optional rigidity mask that allows the user to paint an area whose features are preserved.
}
\tibin{
This novel formulation supports realistic shape deformation of dense meshes through vertex-level editing of a small number of vertices.
More specifically, given a number of vertex constraints, a region of influence in the model, and a very brief text description of the object (could just be the object name, like chair, car, etc.), \tibin{an optional rigidity mask}, our method produces realistic and meaningful deformations with just a few user inputs yielding better results than traditional methods and unlike many earlier works, it is not restricted to any specific class or classes of geometric shapes. We repeat that the optional rigidity mask allows the user to select geometric features that should be preserved during deformation. 
 }

\tibin{
Our main contribution is a direct 3D mesh editing algorithm with user-specified feature preservation that yields a global context-aware realistically deformed mesh by dragging just a few mesh vertices.  More specific contributions are:
}
\tibin{
\begin{itemize}
    \item We integrate the requirement of satisfying user-specified geometric constraints while utilizing the global context obtained through the use of 2D priors from a large-scale generative model. We believe our approach is the first in this direction. 
    \item A new formulation of the mesh deformation problem through a novel geometric regularizer that allows more granularity to the deformation result and more specific user control of stretch and rigidity.
    \item A new loss function is defined as a weighted combination of user-specified vertex constraints, \tibin{a geometric regularizer} and a DDS score using the 2D prior collectively optimizes for geometry and realism in the deformed shape.
\end{itemize}
}

\section{Related work}
\begin{figure*}[htb]
   \centering
   \includegraphics[width=\linewidth]{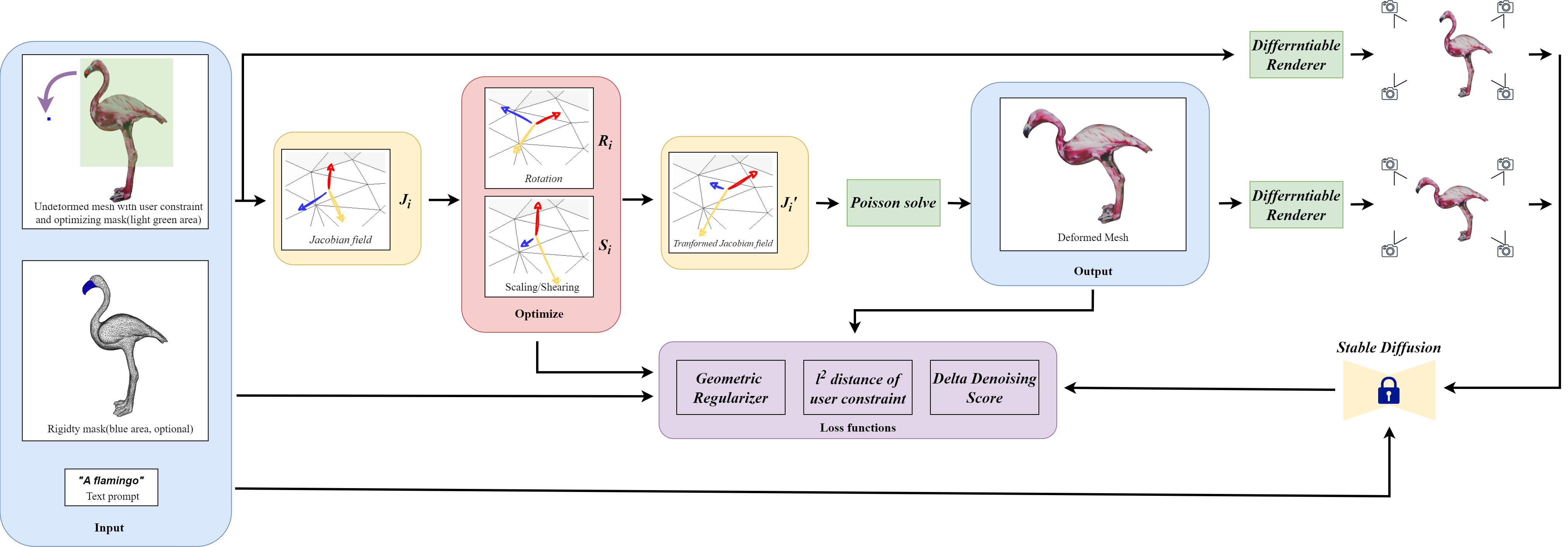}
   \caption{\label{fig:overview}
     Overview of our method. Input: The user specifies constraints (vertices in red disks to be moved to new positions in blue disks) and an optional rigidity mask (painted in blue). Our approach combines these constraints with a shape regularizer and a DDS score (natural image prior) applied on multiple views to perform the mesh deformation. Output: resulting deformed mesh.}
\end{figure*}
In basic terms, geometric editing techniques can be divided into data-driven and non-data-driven methods.
Non-data-driven shape deformation techniques typically frame the editing operation as an optimization problem~\cite{yuan2021revisit} where user constraints are coupled with a regularizer that guides the rest of the shape.
Shape regularizers come in two flavors: physically-based regularizers that try to minimize physically-based energy functions, often based on elastic energies~\cite{chen2023local} and geometric regularizers that try to preserve the original shape of the object and are often based on shape differential operators~\cite{botsch2007linear,sorkine2007rigid,chao2010simple}.

With the introduction of neural representations of geometry Yang et al.~\cite{yang2021geometry} proposed a neural implicit level set representation that supports editing operations, but the edited shapes are obtained using an optimization framework with physically-based losses and not data-driven constraints.
Yuan et al.~\cite{yuan2022nerf,yuan2023interactive} proposed a geometric editing method based on the NERF representation. However, the editing is performed using the classical as-rigid-as-possible (ARAP) method applied on a mesh obtained from the NERF and which is converted back to the NERF representation afterward.

Regardless of representation, one of the shortcomings of these methods is the lack of a high-level semantic understanding of the object; so they often result in unrealistic deformations. 
To address such shortcomings, some papers rely on additional heuristics: ~\cite{kraevoy2008non, gal2009iwires} such as maintaining the proportion of geometric features taking inspiration from CAD design or say, editing the object using a set of salient curves on the object. 
However, a more general solution is to use a data-driven mechanism to guide the deformation.

\subsection{Data-driven Geometric Editing}

Data-driven geometric editing methods are tailored to the representation of the underlying geometry.
Implicit neural representations such as DeepSDF~\cite{park2019deepsdf} or NERF~\cite{mildenhall2021nerf} are popular due to their generative prowess. 
Yumar et al.~\cite{yumer2016learning} proposed a method that uses an implicit occupancy grid to learn the space of shapes from a set of 3D data.
Deng et al.~\cite{deng2021deformed} propose a method that uses a neural signed distance function representation that also learns a neural shape model from a collection of 3D objects.
Both of these methods are limited to a small and specific set of objects.
Liu et al.~\cite{liu2021editing} introduced a method that learns a conditional radiance field over an entire object class to guide the deformation behavior. However, 
all the above methods are subject to a common limitation of being reliant on the availability of large amounts of relevant 3D data which is much less accessible as compared to 2D images.

One way to overcome this limitation is to leverage large-scale image-based models such as CLIP~\cite{radford2021clippaper} or stable diffusion~\cite{poole2022dreamfusion}.
Since models like CLIP and stable diffusion use text prompts as a controlling mechanism, it has opened the door to text-based 3D generation methods~\cite{poole2022dreamfusion,mohammad2022clip}.
For purposes of shape editing, 
Hyung et al.~\cite{hyung2023local} propose a purely text-based editing framework using a NERF representation and 
Mikaeili et al.~\cite{mikaeili2023sked} propose a sketch-based editing method using a NERF representation. 
The latter method is close to ours in spirit in that it uses the SDS loss~\cite{poole2022dreamfusion} to guide the deformation using a sketch-based interface.
However, this method does not allow for direct control over the geometry, an operation that is very challenging when using an implicit representation and would also necessitate user interaction in several views.

Despite the popularity of the neural implicit representations mentioned above, 3D triangular meshes are still very much in use in many real-life applications due to their simplicity, efficiency, and downstream hardware processing support through GPUs.
Wang et al.~\cite{wang20193dn} propose a method that trains an end-to-end network that deforms the source model to resemble the target. 
Because the method infers per-vertex offset displacements, it is not suited for vertex-based mesh editing say, by specifying only a few vertex constraints. 
Wang et al.~\cite{yifan2020neural} propose a neural cage network that infers cage coordinates of the points inside. 
Both these networks are trained by combining shape-based Laplacian losses and other heuristics tailored toward generic man-made objects.
Therefore despite being neural methods, the deformation behaviour learned is not driven directly by 3D or deformation data. 

Early data-driven methods~\cite{sumner2005mesh,popa2007interactive} focus on learning a deformation behavior from a set of sample poses in terms of deformation gradients.
More recently Tang et al.~\cite{tang2022neural} used supervision to learn the prior deformation of a specific class of objects, mostly animals from an existing database.
\tibin{All} of the above approaches rely not only on the availability of 3D data, but also on the availability of sample deformations of the same mesh for supervised learning.
Jakab et al.~\cite{jakab2021keypointdeformer} discover key feature points in a dataset of objects and cast the problem as transforming a source 3D object to a target 3D object from the same object category. 
The feature points used for deformation are not user-selected.
Furthermore, these mesh-based methods have similar limitations to their neural counterparts in that they rely on a 3D database of objects of a certain class.

To overcome this limitation Gao et al.~\cite{gao2023textdeformer} edit a triangular mesh using only a text prompt using CLIP embeddings. 
One major challenge with the CLIP methods is that CLIP embeddings do not encode a viewing direction resulting in ambiguities denoted as the Janus effect~\cite{gao2023textdeformer}. The other major challenge is that it seems difficult to accommodate the satisfaction of user-specified geometric constraints.

The use of large-scale pre-trained models for 2D image editing has been explored in several works. Pan et al.~\cite{pan2023drag} propose a method for point-based image editing by optimizing in the latent space of the StyleGAN2 generator~\cite{karras2020analyzing}. Similarly, Shi et al.~\cite{shi2023dragdiffusion} and Mou et al.~\cite{mou2023dragondiffusion} achieve something similar by optimizing the diffusion latent space. In ~\cite{gao2022get3d}, 3D textured meshes are generated from the learned latent space of images. 
Unfortunately, these methods cannot be extended to direct vertex editing of a given 3D mesh, as they are completely reliant on locating the deformed representation in the rich latent space of 2D images. As already mentioned, we do not have the luxury of such a rich latent space for 3D shapes.  In our work, we also use the latent space of images, but we use it to regularize the 3D geometry by enforcing it to result in a natural image when rendered from an arbitrary view.

In summary, our main goal in this work is to provide a general mesh editing method with user-specified vertex constraints that do not involve supervision, hence does not rely on 3D data for training, and produces realistic results without being restricted to a specific class of objects. 
\tibin{We achieve this by (1) harnessing the rich and vast knowledge about natural and human-made objects that are represented in today's pre-trained large-scale models in the image format and (2) couple it with a flexible geometric regularizer for accurate user control. Our method is described in the following section.}

\begin{figure*}[htb]
   \centering
   \includegraphics[width=0.9\linewidth]{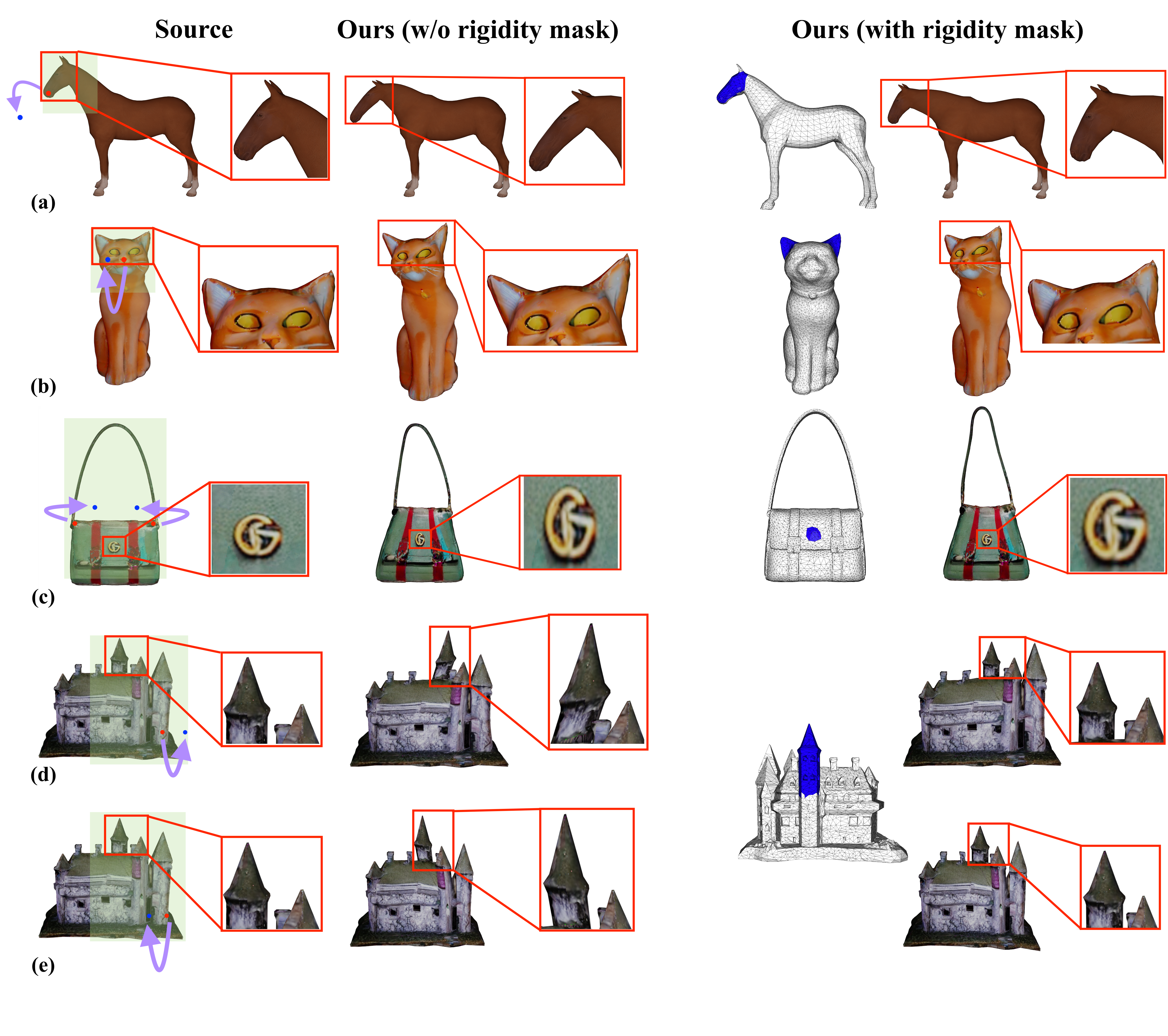}
   \caption{\label{fig:results_rigidity}
     Showcasing applications of rigidity mask (shown in blue)}
\end{figure*}

\section{Method}
The overview of our proposed method is shown in Figure \ref{fig:overview}.
\tibin{
The user specifies: (1) a set of mesh vertices (red disks) as handles to drag, paired with a corresponding set of target 3D positions (blue disks); (2) a region of influence to specify the part of the mesh that is allowed to be modified (rendered as a transparent green rectangle), (3) a short (typically one word) prompt generally describing the object (e.g., a car, a chair, etc.) and (4) an optional rigidity mask (rendered in dark blue) that indicates geometric features that should be preserved during deformation.
}

\tibin{
We optimize directly for the transformation of the Jacobian field represented as an explicit combination of rotation and stretching
for each triangle in the region of influence guided
 by three losses: (1) $l^2$ distance of the user constraint loss, (2) \tibin{a geometric loss to control the local geometric behavior and enforce a smooth deformation and (3) Delta Denoising Score (DDS) loss which makes the deformed mesh have realistic appearances when rendered from random viewpoints and provides global guidance to our 3D model.}
}

\tibin{The remainder of the paper is organized as follows. Section~\ref{jacobians} shows our deformation formulation using the neural jacobian fields, section~\ref{sec:DDS} reviews the Delta Denoising Score, section~\ref{all} explains how everything fits together and provides the editing workflow and additional implementation details. 
Section~\ref{res:comp} presents our experiments and analysis of the method including ablation studies related to our design decisions, and finally, Section~\ref{conclusion} has conclusions, limitations, and future work.
}
\subsection{Transformations of the Neural Jacobian field}
\label{jacobians}
The neural Jacobian field \cite{aigerman2022neural} operates in the intrinsic gradient domain of triangular meshes to learn highly accurate mappings between meshes. 
For triangular mesh vertices $\Phi$, the per-face Jacobian $J_i \in \Re^{3\times3} $ is defined as 
\begin{equation}
    J_i = \Phi \nabla^T_i,
\end{equation}
where $\nabla^T_i$ is the gradient operator of triangle $t_i$. Given matrices $M_i \in \Re^{3\times3}$ for every triangle, we can compute vertex positions $\Phi^*$ whose Jacobians $J_i=\Phi^*\nabla^T_i$ are least-square closest to $M_i$ by solving the Poisson equation. The solution is obtained by solving the following linear system
\begin{equation}
    \Phi^* = \mathcal{L}^{-1}\mathcal{A}\nabla^TM,
\end{equation}
where $\mathcal{A}$ is the mesh's mass matrix, $\mathcal{L}$ is the mesh's Laplacian, and $M$ is the stacking of the input matrices $M_i$. (For a detailed mathematical definition of the Jacobian field, please refer to \cite{aigerman2022neural}). 

\tibin{
 Optimizing the mesh by optimizing the Jacobian field can produce a more smoothly-deforming mesh and avoid entanglement as compared to optimizing the mesh directly~\cite{gao2023textdeformer} as we illustrate in Figure~\ref{fig:ablation}.
 Suppose the initial per-face Jacobian is $J_i \in \Re^{3\times3}$ for face $i$, and the deformed per-face Jacobian is $J_i^{'}$. There is a linear transformation $T \in \Re^{3\times3}$ that $J_i^{'} = TJ_i$. By polar decomposition, the transformation matrix can be decomposed to 
 \begin{equation}
     T = R*S,
 \end{equation}, where $R$ is the rotational component (orthogonal matrix) and $S$ is the stretching component (symmetric semi-positive definite matrix) of the transformation.
 Thus, we reverse the process and we express our deformation as a per-triangle rotation and stretching in the Jacobian space. We represent the rotational component using the smooth 6-DoF representation illustrated in \cite{zhou2019continuity} and the scale/shear (stretch) component as a symmetric 3 by 3 matrix with 6 degrees of freedom.
 Now we can define a geometric regularizer to produce a smooth deformation directly on the variables as: 
 \begin{equation}
     \mathcal{L}_{reg}(i) =  \Sigma_{j\in N}(a_i \| R_i-R_j\|_2^2+b_i \|S_i-S_j\|_2^2) + c_i \|S_i-I\|_2^2 + d_i \|R_i-I\|_2^2 ,
     \label{eq:geo_reg}
 \end{equation}
 where $N$ is the set of adjacent faces of face $i$ and $I$ is the identity matrix. It should be noted that by adjusting the weights in equation \ref{eq:geo_reg}, we can control the per-face rigidity of the deformation. 
 Details on setting these parameters are provided in section ~\ref{impldetails}
 }
 
To keep the smoothness of the boundary between the influenced area and the other area, a boundary regularizer needs to be applied,
\begin{equation}
    \mathcal{L}_{bd} = \Sigma (\|V_{bd}^{'} - V_{bd}\|^2_2) + \Sigma(\|R_{bd}-I\|^2_2 + \| S_{bd}-I\| ^2_2 ),
\end{equation}
where $V_{bd}$ are the boundary vertices of the original mesh, and  $V_{bd}^{'}$ are the boundary vertices of the deformed mesh, $R_{bd}$ and $S_{bd}$ are the rotational and stretching matrices of the boundary faces, respectively.


\subsection{Delta Denoising Score}
\label{sec:DDS}
Score Distillation Sampling(SDS) was introduced in \cite{poole2022dreamfusion} to use a $2D$ prior to guide the synthesis of $3D$ shapes. Delta Denoising Score(DDS)\cite{hertz2023delta} extended the $SDS$ mechanism for guidance in image editing.
Given an input image $x$, a text embedding $z$, a denoising model $\epsilon_{\phi}$ with parameters $\phi$, a randomly sampled time step $t \sim \mathcal{U}(0,1)$ drawn from the uniform distribution, and noise $\epsilon \sim \mathcal{N}(0,I)$ following a normal distribution, the weighted denoising score can be expressed as
\begin{equation}
    \mathcal{L}_{Diff}(\phi,x,z,\epsilon,t) = w(t) \| \epsilon_{\phi}(x_t,z,t)-\epsilon \|^2_2
\end{equation}
where $w(t)$ is a weighting function that depends on the time step $t$, and $x_t$ is the $x$ added the noise at time step $t$. For text-conditioned diffusion models that use classifier-free guidance~\cite{ho2022classifier}, the denoised image is expressed as  
\begin{equation}
    \hat{\epsilon}_{\phi}(x_t,z,t) = (1+w)\epsilon_{\phi}(x_t,z,t)-w\epsilon_{\phi}(x_t,t),
\end{equation}
which is a weighted sum of conditioned and unconditioned denoising.

Thus, as shown in \cite{poole2022dreamfusion}, given an arbitrary differentiable parametric function $g_{\theta}$ that renders images, the gradient of the $SDS$ guidance is given by

\begin{equation}
    \nabla_{\theta} \mathcal{L}_{SDS}(x,z,\epsilon,t) =  \hat{\epsilon}_{\phi}((x_t,z,t)-\epsilon){{\partial x_t}\over {\partial\theta}} .
\end{equation}

Using that gradient to optimize the $g_{\theta}$ can produce images that look natural. However, for image editing, \cite{hertz2023delta} has shown that $SDS$ can produce non-detailed and blurry outputs due to the noisy gradient. To overcome this problem, $DDS$ was introduced which does two $SDS$ processes for the edited image and reference image respectively, and retrieves the gradient by subtraction of these two,
\begin{equation}
    \label{eq:DDS}
    \nabla_{\theta} \mathcal{L}_{DDS} = \nabla_{\theta} \mathcal{L}_{SDS}(x_{edit},z_{edit}) - \nabla_{\theta} \mathcal{L}_{SDS}(x_{ref},z_{ref}),
\end{equation}
where $x_{edit}$ and $x_{ref}$ are the edited image and reference image, $z_{edit}$ and $z_{ref}$ are the text prompts of the edited image and reference image. It shows that the $DDS$ has a less noisy gradient and can produce a higher fidelity image in the image editing task.

\subsection{Putting it all together: DragD3D}
\label{all}

Suppose a user constraint start point is $c_i$, and target point is $c_i^{'}$, then the user constraint loss is given by
\begin{equation}
    \mathcal{L}_{user} = \sum_{i=0}^N \|c_i^{'}-c_i\|_2^2,
\end{equation}
where $N$ is the number of user-specified handles. 
The $DDS$ gradient is computed as per equation \ref{eq:DDS}. In our case, suppose the undeformed mesh is $\Phi$, the deformed mesh is $\hat{\Phi}$, the random viewpoint is $vp$, the differentiable renderer is $g(\cdot)$, and the text prompt is $z$, the $DDS$ score of the deformed mesh can be expressed as
\begin{equation}
    \nabla_{\hat{\Phi}} \mathcal{L}_{DDS} = \nabla_{\hat{\Phi}} \mathcal{L}_{SDS}(g(\hat{\Phi},vp),z)- \nabla_{\Phi} \mathcal{L}_{SDS}(g(\Phi,vp),z).
\end{equation}

For the denoising model $\epsilon_{\phi}$, we use the Stable Diffusion\cite{rombach2022sd}.
Overall, we optimize for 
\begin{equation}
    \label{total_loss}
    \mathcal{L}_{total} = \lambda_{user} \mathcal{L}_{user} + \lambda_{DDS} \mathcal{L}_{DDS} + \mathcal{L}_{reg} + \lambda_{bd}\mathcal{L}_{bd},
\end{equation}
where the weight of the user constraint $\lambda_{user}$ was set to $50$, and $\lambda_{bd}$ was set to $100$.

For DDS guidance, we use Stable Diffusion 2.1\cite{rombach2022sd} as the backbone along with the $Perp-Neg$ algorithm~\cite{perp_neg}. The guidance scale is $100$, and the gradient scale is $0.00002$.

\subsection{Controlling the deformation result}
\tibin{
User controllability of the deformation is a key requirement of any deformation method. 
In addition to the user-selected anchors, we provide two additional optional mechanisms for guiding the deformation.
The first is the use of a rigidity mask: an area of the object easily selected by the user with the purpose of inhibiting local deformations in the area. This can be useful in situations such as the ones illustrated in Figures~\ref{fig:results_rigidity} to ensure that meshes preserve their features instead of simply deforming (i.e., keep the same horse when moving its nose/head, instead of deforming into a different horse that has its nose/head in the right place). If a rigidity mask is applied, the weights in the Loss function in equation~\ref{eq:geo_reg} are changed to add more weight to the variables corresponding to the triangles in the masked area.

The second mechanism is inhibiting the rotations. In some examples, typically of man-made objects such as the Lantern in Figure~\ref{fig:teaser} the intent of the user is likely only to stretch the object and not bend it. 
Allowing both rotations and stretching components will tend to also add some undesired bending to the result. 
In our formulation, we can inhibit the rotational component thus allowing only stretching which will result in a more natural deformation.
}

\subsection{Implementation details}
\label{impldetails}

\tibin{
For all examples in this paper, we optimize for $2000$ iterations. 
We use Adan\cite{xie2022adan} as the optimizer with a learning rate set to $0.005$. 
For example, large $b_i$ and $c_i$ will constrain the face stretching and large $a_i$ and $d_i$ will constrain the face rotation. 
}
\tibin{
Normally, when the deformation type is not selected, we set the following default values for the weights: $a_i = 0.6$, $b_i = 0.12$, and $c_i = d_i =0$. 
For deformations where we want to inhibit rotations, we use $a_i = 100$, $b_i = 1$, $c_i = 0$ ,and $d_i = 100$. 
For the faces of the rigidity mask, we set  $b_i=100$ and $c_i = 100$.
}
\tibin{
As noted in~\cite{mohammad2022clip} such a large-scale optimization problem can easily get stuck in an undesired local minimum. We minimize this possibility by doing the following to help the optimization process. 
In every iteration, four camera positions were sampled randomly with azimuth $ \in [-180^ {\circ}, 180^{ \circ}]$, elevation angle $ \in [0^ {\circ},90^ {\circ}]$, camera distance$ \in [d_0,d_0+2]$, where $d_0$ is the camera distance of specifying handles. 
}

\section{Experiments and Discussion}
\label{res:comp}
\subsection{Results}
\tibin{
We demonstrate the effectiveness and generality of our method on a variety of meshes belonging to different object classes, engineering, and free-form shaped objects. 
A gallery of results is shown in Figures~\ref{fig:teaser},\ref{fig:results_rigidity}, \ref{fig:results1}, and ~\ref{fig:results2} as well as the accompanying video.
All the meshes were rendered with and without texture to show the geometry clearly. 
The meshes were obtained from TurboSquid, Thingi10K\cite{zhou2016thingi10k}, and TEXTure\cite{richardson2023texture}. 
For the meshes without texture (car, lantern, flamingo, cat, handbag, castle, dragon, rocket, chair), we generated the texture using \cite{richardson2023texture}.
In figures \ref{fig:results1} and ~\ref{fig:results2}, the first column shows: the original image, the region of influence, and the user constraints; the second column shows our result, including highlighting certain features or rendering from an alternative viewpoint and the third column shows a result generated using the ARAP~\cite{chao2010simple} method. 
As can be seen from all the experiments, our deformation results are realistic, smooth and satisfy the user constraints.
}

\tibin{
In figures \ref{fig:results1} and ~\ref{fig:results2} the results are generated using the default weights except for the lantern and castle. For the lantern example, we use the weights specified in section~\ref{impldetails} to inhibit the rotations and for the castle example we simultaneously use a mask to avoid a shearing artifact in one of the castle's columns and also we inhibited the rotations.
}

\subsection{Rigidity Mask}
\tibin{
Figures ~\ref{fig:teaser} and ~\ref{fig:results_rigidity} show several examples that highlight the impact of the rigidity mask. 
In the flamingo (Figure~\ref{fig:teaser}c) and horse (Figure~\ref{fig:results_rigidity}a) examples, the default deformation distorts the beak and head of the object, respectively. 
Increasing the deformation weights of the stretch component for triangles inside the rigidity mask will preserve the original geometric features yielding more realistic results. 
Furthermore, in the example of the cat (Figure~\ref{fig:results_rigidity}b), when we rotate the head, the ears of the cat change shape. The ears of the "new" cat still look realistic, but the identity of the cat is lost as they are not the same ear as before. By adding the ear regions to the rigidity mask we can obtain a result that preserves the identity of the cat.
}

\tibin{
In the castle examples (Figure~\ref{fig:results_rigidity} d,e), when using the default values one of the columns presents with a shearing artifact that can be fixed by painting a rigidity mask over the column thus keeping the geometric structure of the object intact.
Lastly, in the deformation of the bag (Figure~\ref{fig:results_rigidity} d,e), the logo unsurprisingly gets distorted and adding the rigidity mask around the logo will better preserve the logo shape.
}

\subsection{Comparison with optimization-based methods}
\tibin{
Our main goal is to provide a general mesh editing method that provides realistic results by leveraging the large-scale 2D models developed with millions of images of vast classes of objects from multiple views. Unlike many other data-driven methods, our method does not need data-driven guidance and is not constrained to a specific class of objects or based on a given set of sample deformations. 
Therefore, our comparison is with a method that does not use any data-driven guidance for the deformations.
While there is a multitude of such methods (we refer to two state-of-the-art reviews~\cite{botsch2007linear,yuan2021revisit})
for practical reasons (i.e. nonavailability of code, no clear winner in terms of deformation quality), from this multitude of methods we pick the as rigid as possible as a representative of the optimization-based class of methods~\cite{yuan2021revisit}.
}
\tibin{
}

\begin{figure*}[htb]
   \centering
   \includegraphics[width=\linewidth]{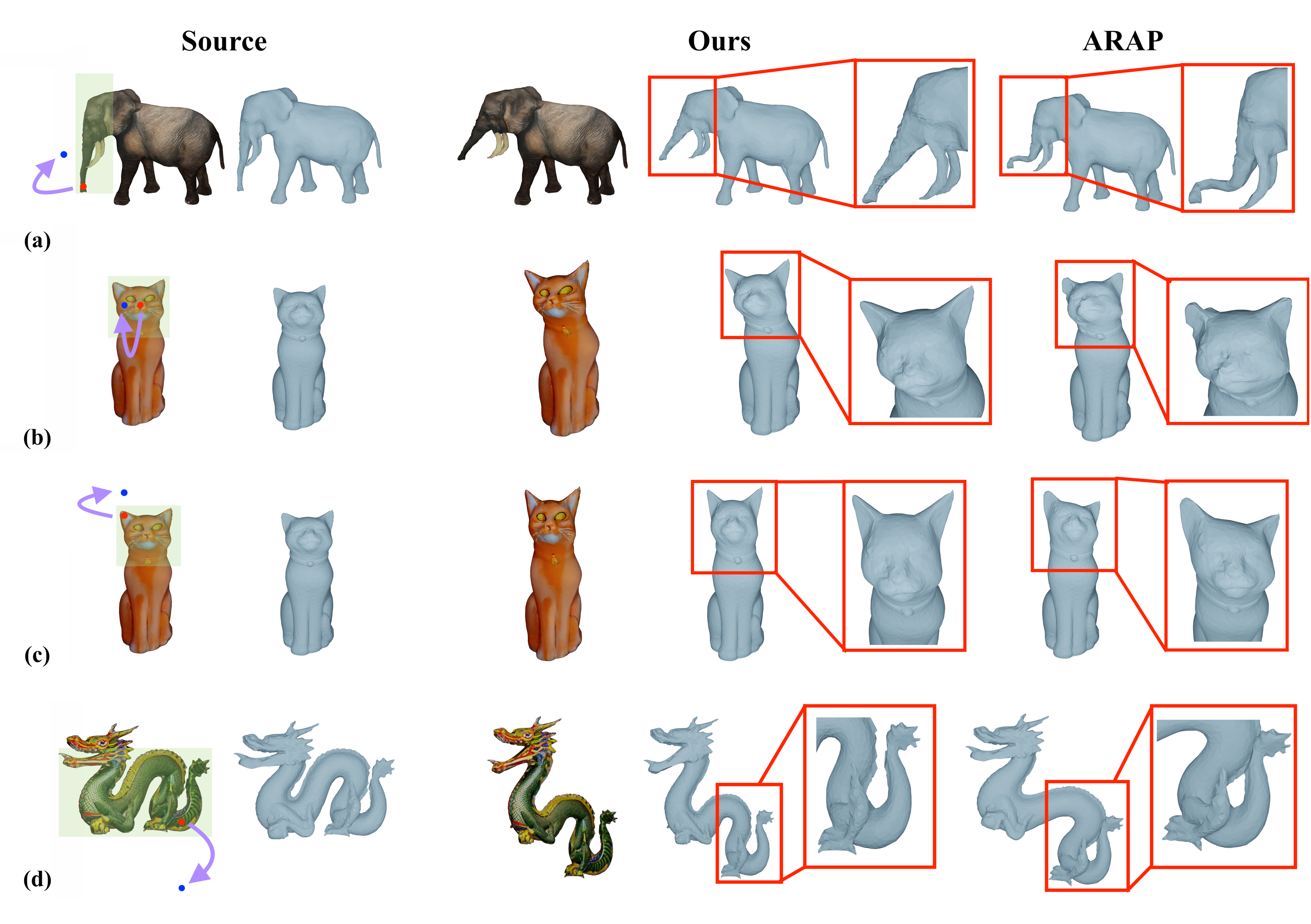}
   \caption{\label{fig:results2}
     Gallery of results 2 (free-form objects)}
\end{figure*}

Although ARAP is an effective and widely used geometry deformation method, its lack of global contextualization yields some less natural results compared to our method as shown in Figures ~\ref{fig:teaser}, \ref{fig:results1} and \ref{fig:results2}, both in terms of global and local feature preservation.

\textbf{Global structure preservation:} The large-scale image model is also aware of the global structure of objects including symmetry and postures. As shown in Figure~\ref{fig:teaser}$(a)$ and \ref{fig:results1}$(d)$, the roof of the car was lifted on both sides preserving the symmetry of the object. 
In Figure~\ref{fig:results2} $(c)$, although only one ear of the cat was pulled up, the other ear was also deformed to keep the symmetry of the two ears. However, if the user desires the change in only one ear, then a rigidity mask would address this need. 
In Figure~\ref{fig:results2} $(a)$, rather than only deforming the trunk, the face of the elephant was also slightly lifted, making the deformation look more realistic; 
In Figure~\ref{fig:results2}$(d)$, the middle part of the dragon was also deformed to avoid the entanglement between the tail and body, as can be seen in the ARAP result.

\textbf{Local feature preservation:} The large-scale image model that we use is also aware of the local features of objects including flat surface and smooth curvature even away from the anchors. Thus, unlike ARAP, our results preserve implicitly the features of deformed objects. As shown in Figure~\ref{fig:teaser} $(a)$, the car body near the rear window was deformed smoothly by our method, but ARAP produced a hump at nearly right angle; Figure~\ref{fig:teaser}$(b)$ the bottom of the lantern was kept flat by our method, but ARAP made a sharp artifact. 
In figure~\ref{fig:results1}$(b)$, the head of the cat is well preserved while the ARAP result the head is distorted.
In Figure \ref{fig:results2}$g$, the top of the chair was deformed to a headrest that has a small arc, and the armrest was kept horizontal.

\begin{figure}[tb]
    \centering
    \includegraphics[width=\linewidth]{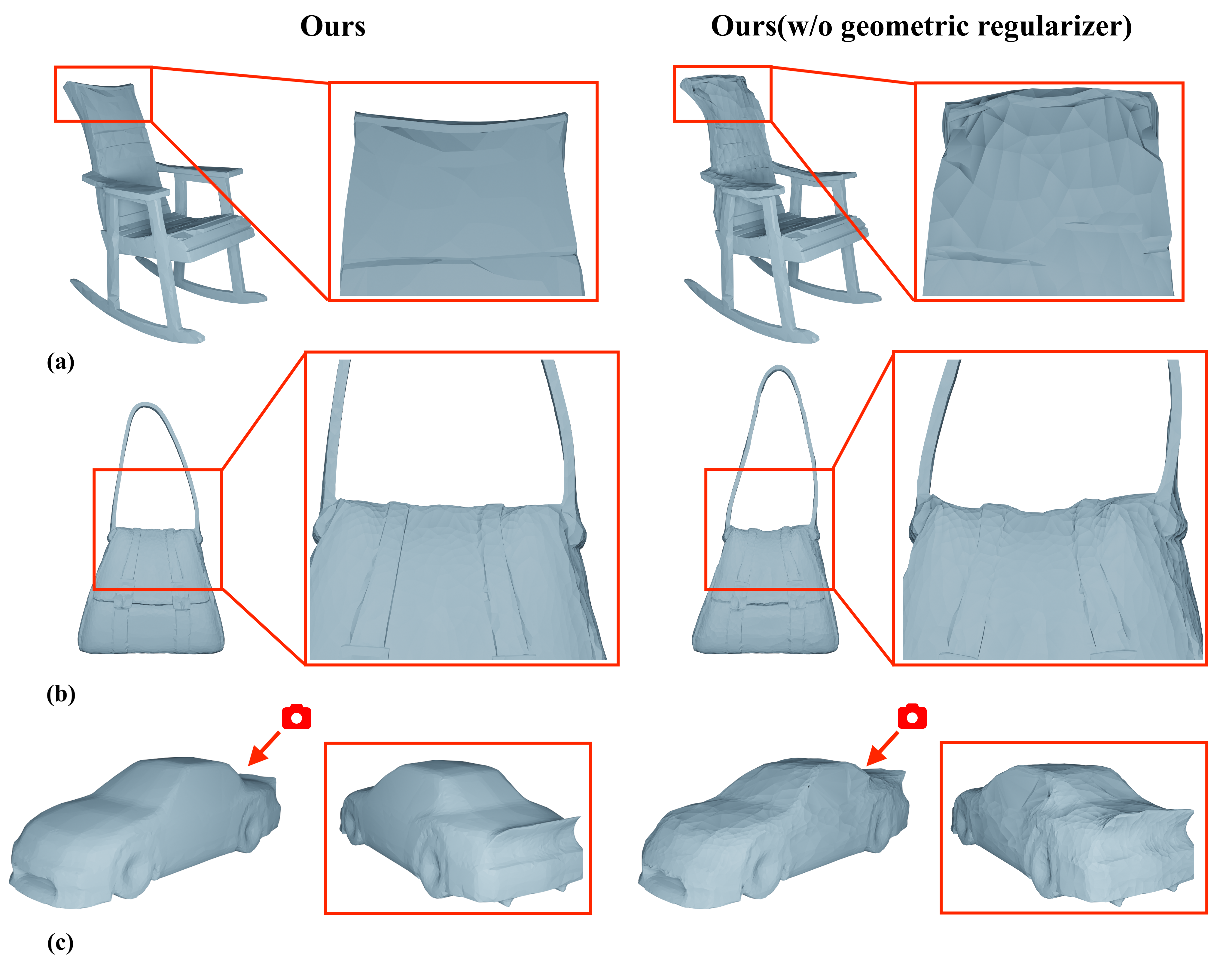}
    \caption{Ablation study showing the effectiveness of geometric regularizer.}
    \label{fig:ablation_reg}
\end{figure}
\subsection{Ablation Studies}

\begin{figure*}[htb]
    \centering
    \includegraphics[width=\linewidth]{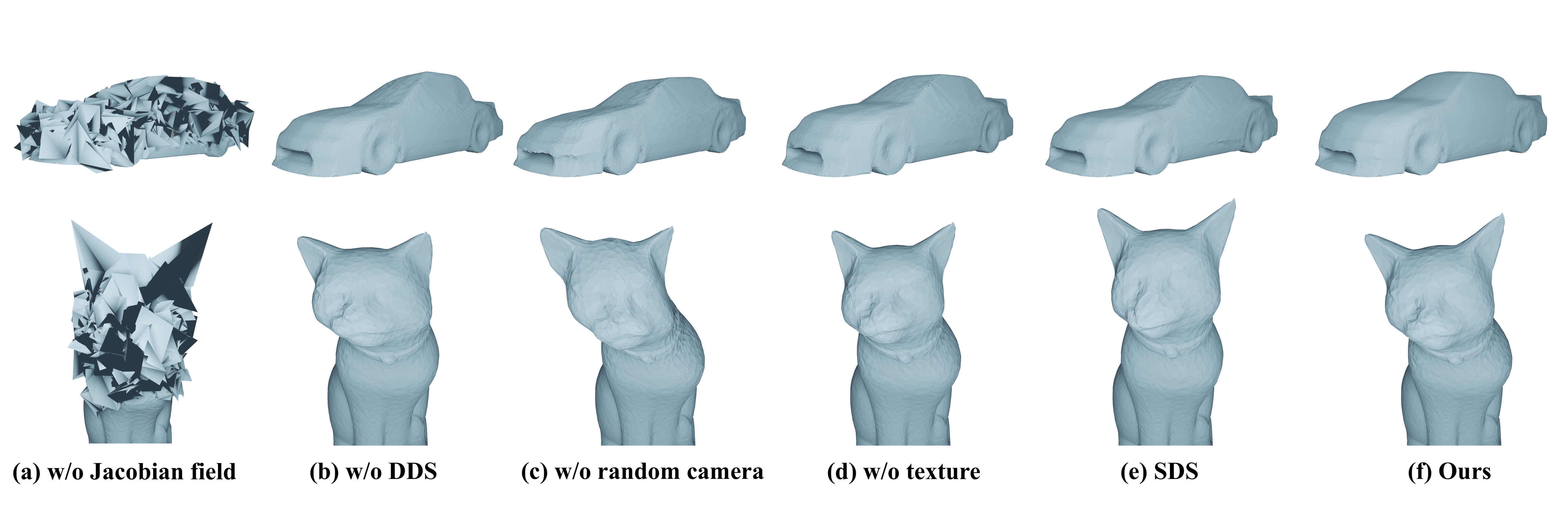}
    \caption{\tibin{Ablation study. $(a-d)$: removing individual features of our method. $(e)$: using SDS instead of DDS. $(f)$: our method}}
    \label{fig:ablation}
\end{figure*}

We perform various experiments to show the effectiveness of the Jacobian fields, geometric regularizer, random camera views, guidance, and texture. As can be seen in Figure \ref{fig:ablation}, if we optimize for vertex displacement directly, the mesh structure cannot be kept.  If the camera was fixed on four canonical views, front, back, and sides, and without any change in the elevation angle, the deformation was overfitted to these four views, which produced non-smooth deformation.  As shown in Figure \ref{fig:ablation}, if we replace the $DDS$ guidance with the $SDS$ guidance in our pipeline, due to the noisy gradient of $SDS$, the edited mesh is less smooth and has more distortion. Without texture, the guidance of the 2D diffusion priors is not as efficient as before, which makes for lower-quality editing. As shown in figure \ref{fig:ablation_reg}, without the geometric regularizer, the mesh is not as smooth as we include geometric regularizer which leads to some artifacts, for example, in $(c)$ the anchor near the rear window produced a sharp artifact. Local features are also not preserved well, such as in Figure \ref{fig:ablation_reg} $(b)$, the top of the bag is not flat enough when the geometric regularizer is not applied.

\begin{figure}[t]
    \centering
    \includegraphics[width=0.8\linewidth]{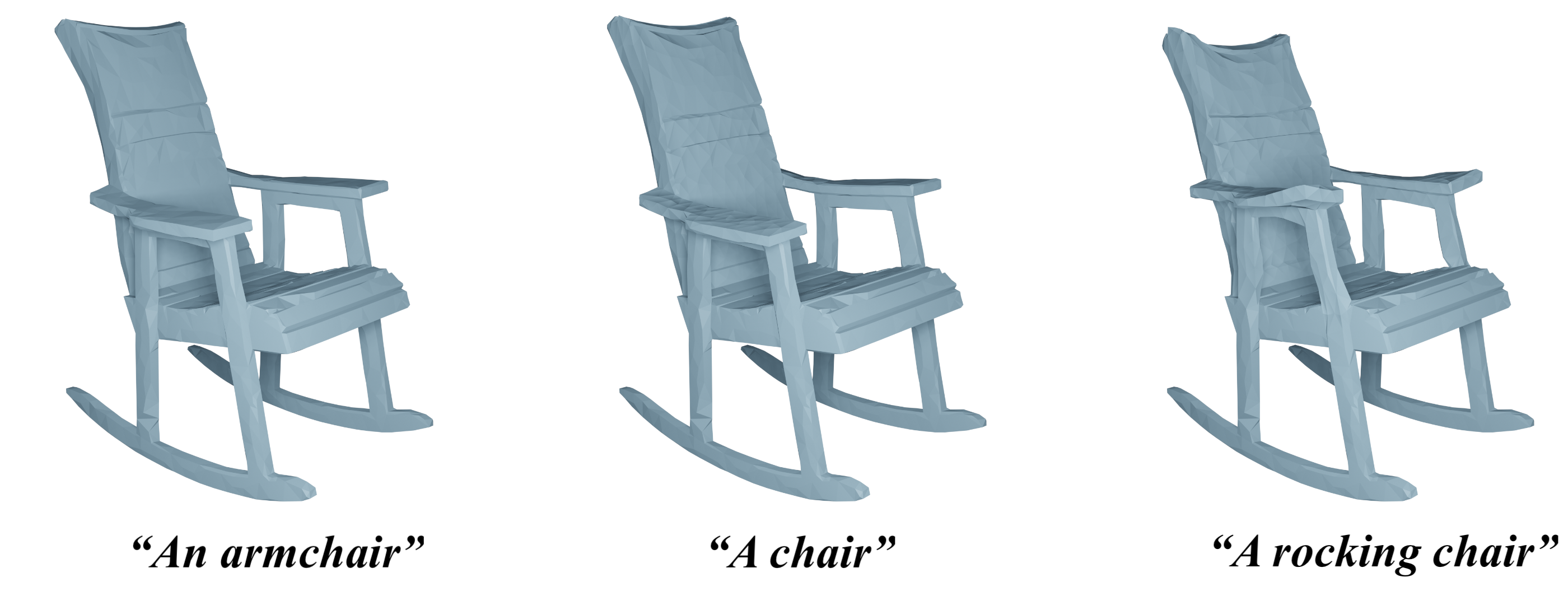}
    \caption{Same deformation with different prompts. The prompts need not be very precise.}
    \label{fig:ablation_chair}
\end{figure}

Lastly, we test dependence on the precision of the text prompt given by the user. As Figure \ref{fig:ablation_chair} shows, the description does not need to be very precise, as long as it matches the object. This is clear as the prompt is used only in the DDS loss and primarily serves to provide the global context for this object's shape.



\section{Conclusions, Limitations, and Future work}
\label{conclusion}

Dragging 3D mesh vertices in space has been used in practice to provide the fine control that designers seek in shape design. For dense meshes, this poses the problem of how to automatically determine the change in all other vertices. 
\tibilm{Our deformation method leverages these large-scale models while providing the user with simple and intuitive shape control for natural and smooth-looking results. In this work, we unravel interesting experimental insights on how 2D large-scale models can be used to learn local and global 3D geometric features.
 }


Some limitations of our method along with potential extensions are:
As with most mesh-based methods, the quality of the triangulation can affect the result.
Our deformation method requires a prompt, albeit a simple prompt to accompany the rest of the constraints. This is easy to provide and in the future, we could use automatic prompting~\cite{luo2023scalable3dcaption}.
Diffusion model inference time can be relatively slow. There are emerging 2D generative models that are much faster. So in future work, we will look at ways to incorporate these advances in diffusion model inference speeds.


\bibliographystyle{ACM-Reference-Format}
\bibliography{sample-base}

\begin{figure*}[htb]
   \centering
   \includegraphics[width=\linewidth]{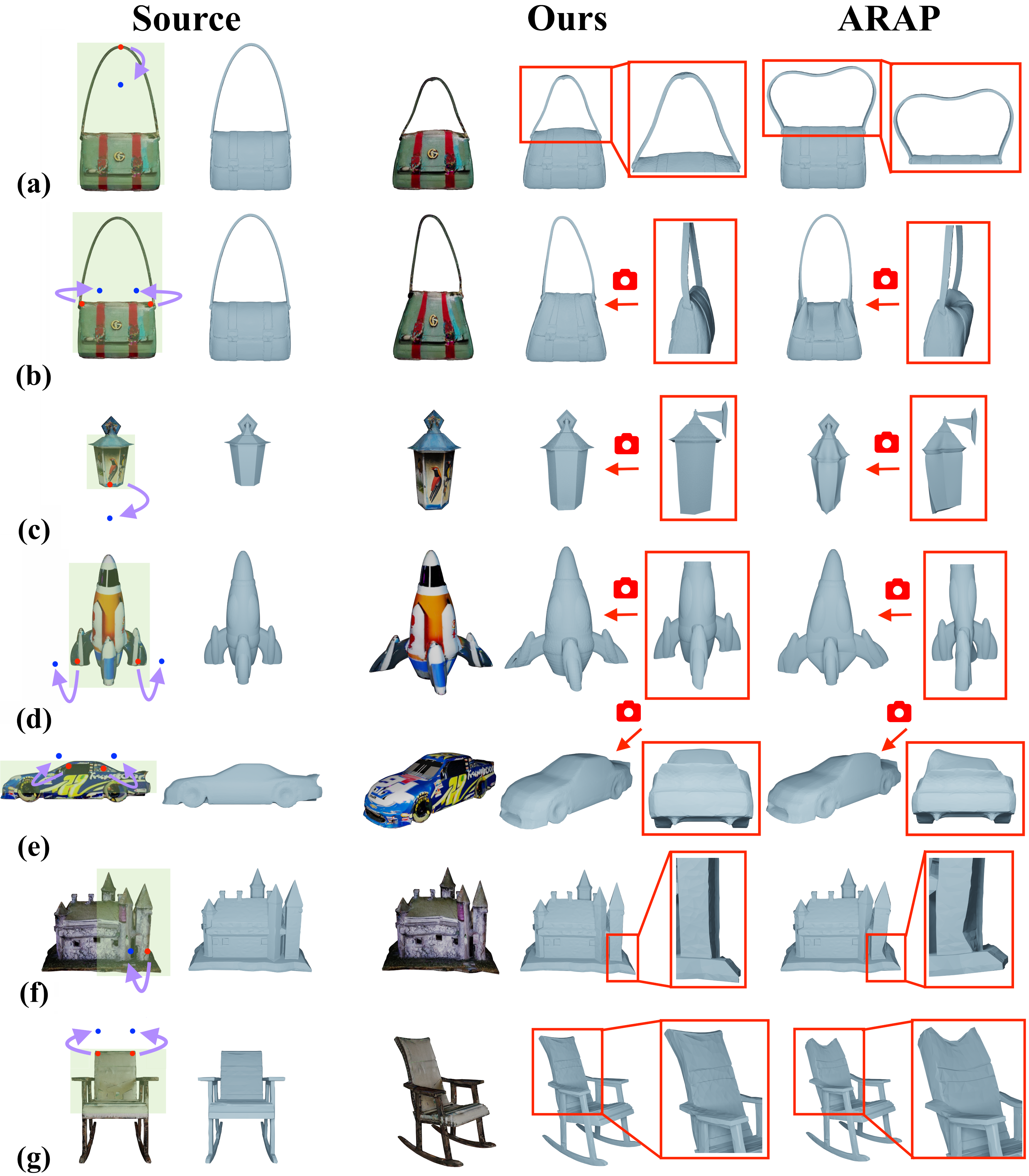}
   \caption{\label{fig:results1_low}
     Gallery of results 1 (engineering objects)}
\end{figure*}

\end{document}